# Impact parameter dependence of isospin effects on the mass dependence of balance energy

Sakshi Gautam[1] and Aman D. Sood[2]
[1]*Department of Physics, Panjab University, Chandigarh-160014, India*
[2] *SUBATECH, Laboratoire de Physique Subatomique et des Technologies Associées, Université de Nantes –IN2P3/CNRS –EMN 4 rue Alfred Kastler, F-44072 Nantes, France.*

The investigation of system size effects in various phenomena of heavy-ion collisions [1] has attracted a lot of attention. The rapid progress in producing energetic radioactive beams has offered an excellent opportunity to investigate various isospin effects in the dynamics of nuclear reaction. Colliding geometry also has a significant role to play in isospin effects [2]. We aim to study the role of colliding geometry in isospin effects on the mass dependence of balance energy. The present study is carried out within the framework of IQMD model [3].

To see the isospin effects, we take two sets of isobars with N/Z =1 and 1.4 throughout the mass range between 48 and 270 covering whole range of colliding geometry from central to peripheral one. We divide impact parameter into four bins i.e. $0.15<b/b_{max}<0.25$, $0.35<b/b_{max}<0.45$, $0.55<b/b_{max}<0.65$, and $0.75<b/b_{max}<0.85$. Interestingly, we see that throughout the mass range, more neutron-rich system has a higher balance energy, which can be attributed to the more neutron-proton cross section as compared to neutron-neutron or proton-proton cross section [2]. The calculated balance energies fall on a line which is of power law nature (proportional to $A^{\tau}$). We see that $\tau$ has a different values for set having N/Z = 1 and 1.4 in the whole range of colliding geometry. The different values of $\tau$ for two sets of isobars is due to the increasing role of Coulomb repulsion in the case of N/Z = 1. If we see the difference in $\tau$ values for set having N/Z = 1 and 1.4, we find that the difference increases with impact parameter. This is due to the more pronounced role of Coulomb potential at peripheral geometries leading to enhancement of isospin effects at peripheral colliding geometries

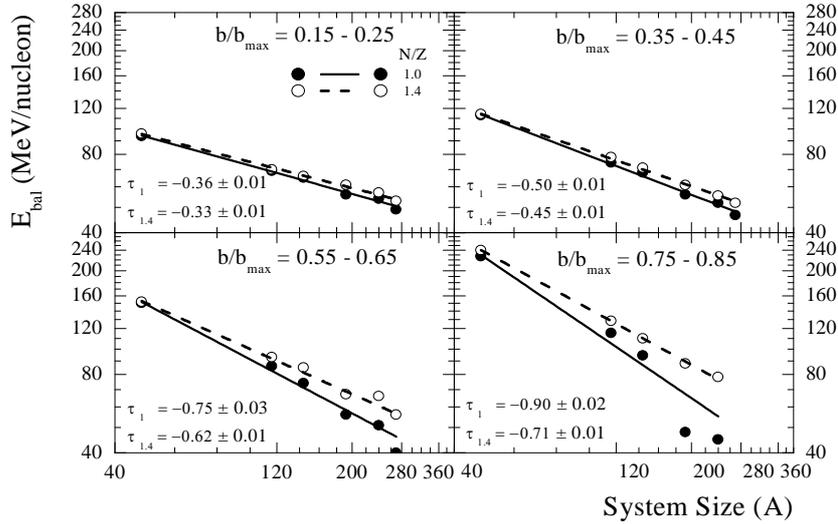

Fig. 1. Balance energy ($E_{bal}$) as a function of combined mass of system in four impact parameter bins. Solid (open) symbols are for systems having N/Z = 1(1.4).